\documentclass[aps, prd, floatfix, twocolumn, superscriptaddress, nofootinbib]{revtex4-1}

\usepackage{latexsym}
\usepackage{amsmath}
\usepackage{amssymb}
\usepackage{amsfonts}

\usepackage{slashed}
\usepackage{bm}

\usepackage[vcentermath]{youngtab}

\usepackage{color}
\definecolor{purple}{rgb}{0.5,0,0.5}
\definecolor{blue}{rgb}{0.0,0,0.9}
\definecolor{prdblue}{rgb}{0.133,0.118,0.498}

\usepackage{enumitem}
\usepackage{multirow}
\usepackage{subfigure}
\usepackage{textcomp}

\usepackage{supertabular} 
\usepackage{placeins}
\usepackage{epsfig}
\usepackage{graphicx}

\begin{document}


\title{The $D_{s0}(2590)^+$ as the dressed $c\bar s(2^1S_0)$ meson in a coupled-channels calculation}

\author{Pablo G. Ortega}
\email[]{pgortega@usal.es}
\affiliation{Departamento de Física Fundamental, Universidad de Salamanca, E-37008 Salamanca, Spain}
\affiliation{Instituto Universitario de F\'isica 
Fundamental y Matem\'aticas (IUFFyM), Universidad de Salamanca, E-37008 Salamanca, Spain}

\author{Jorge Segovia}
\email[]{jsegovia@upo.es}
\affiliation{Departamento de Sistemas F\'isicos, Qu\'imicos y Naturales, \\ Universidad Pablo de Olavide, E-41013 Sevilla, Spain}

\author{David R. Entem}
\email[]{entem@usal.es}
\affiliation{Instituto Universitario de F\'isica 
Fundamental y Matem\'aticas (IUFFyM), Universidad de Salamanca, E-37008 Salamanca, Spain}
\affiliation{Grupo de F\'isica Nuclear, Universidad de Salamanca, E-37008 Salamanca, Spain}

\author{Francisco Fern\'andez}
\email[]{fdz@usal.es}
\affiliation{Instituto Universitario de F\'isica 
Fundamental y Matem\'aticas (IUFFyM), Universidad de Salamanca, E-37008 Salamanca, Spain}
\affiliation{Grupo de F\'isica Nuclear, Universidad de Salamanca, E-37008 
Salamanca, Spain}

\date{\today}

\begin{abstract}
The recent discovery of the $D_{s0}(2590)^+$ meson by the LHCb Collaboration
has stimulated the analysis of meson-meson channels effects in the two-body quark-antiquark meson spectrum.
This resonance, assigned to the radial excitation of the pseudoscalar $D_s^+$ meson,
has a mass much lower than the predictions of naive quark models, which could indicate
a non-negligible $D^{(*)}K^{(*)}$ coupling which reduces its mass.
Based on the importance of nearby meson-meson thresholds in the dynamics of P-wave $D_s$ mesons such as the $D_{s0}^*(2317)^+$ and $D_{s1}(2460)^+$,
in this work we perform a coupled-channels calculation including the $D^{(*)}K^{(*)}$, $D_s^{(*)}\omega$ and $D_s^*\eta$ channels, and study the impact of incorporating those channels
in the mass of the bare $c\bar s$. The coupling between two and four-quark sectors is
done by means of the $^3P_0$ mechanism, with all the parameters constrained from previous studies of the heavy meson spectroscopy.
The masses, widths and production line shapes of the resulting state are analyzed.

\end{abstract}



\maketitle


\section{Introduction}
\label{sec:Introduction}

The heavy-strange spectrum is an excellent system to explore the strong interaction.
In particular, it helps to test and improve low-energy models of quantum chromodynamics (QCD), such
as the quark model.
The charm-strange meson spectrum has been on the spotlight since the discoveries, in 2003, of the
$D_{s0}^*(2317)^+$~\cite{Aubert:2003fg} and the $D_{s1}(2460)^+$~\cite{Besson:2003cp} resonances, two positive-parity states
with unexpected low masses and narrow widths compared to those predicted in quark models for a simple $c\bar s$~\cite{Godfrey:1986wj,Zeng:1994vj, Gupta:1994mw}.

Such discrepancy led to many theoretical interpretations of their inner structure, ranging
from conventional $c\bar{s}$
states~\cite{Fayyazuddin:2003aa, Sadzikowski:2003jy, Lakhina:2006fy} to
molecular or compact tetraquark interpretations~\cite{Barnes:2003dj,
Lipkin:2003zk, Bicudo:2004dx, Szczepaniak:2003vy, Browder:2003fk,
Nussinov:2003uj, Dmitrasinovic:2005gc,Yang:2021tvc}.
In Ref.~\cite{Ortega:2016mms}, these special features of the low-lying $P$-wave $D_s$ mesons were properly
described when meson-meson degrees of freedom are incorporated in the quark-antiquark dynamics. Indeed, the coupling of the $0^{+}$ $(1^{+})$ meson sector to
the $DK$ $(D^{\ast}K)$ threshold was found to be essential to both lower the masses of the bare $c\bar s$ mesons predicted by the naive quark model to those of the corresponding $D_{s0}^{\ast}(2317)$ and $D_{s1}(2460)$ states and describe the $D_{s1}(2536)$ meson as the $1^{+}$ state of the $j_{q}^{P}=3/2^{+}$ doublet, as predicted by heavy quark symmetry, reproducing its strong decay properties.

Recently, the LHCb Collaboration has announced the discovery of a new $J^P=0^-$ $D_s$ meson~\cite{LHCb:2020gnv}, dubbed $D_{s0}(2590)^+$, which has been proposed as the radial excitation of the pseudoscalar $D_s^+$ meson.
Its mass and width are

\begin{align}
 M_{\rm exp}=& 2591\pm 6\pm 7\,{\rm MeV}\nonumber \\
 \Gamma_{\rm exp}=& 89\pm 16\pm 12\,{\rm MeV}.
\end{align}
which are found to be much lower than the masses predicted by naive quark models~\cite{Godfrey:1985xj}. In Ref.~\cite{Xie:2021dwe}, the authors performed
a coupled-channels calculation of the $2^1S_0$ $c\bar s$ and $D^*K$ channel, using the $^3P_0$ mechanism with the aim to reduce the mass of the bare $2^1S_0$ close to the experimental $D_{s0}(2590)$. Neglecting the interaction between the $D^*$ and $K$ mesons, they obtained a renormalized state at $2585$ MeV/c$^2$ with a width of $23$ MeV. The study suggests that the coupling with P-wave meson-meson channels may be relevant to explain the $D_{s0}(2590)$ resonance.

In this letter, we explore the effect of coupling the naive $c\bar s$ $2^1S_0$ to all the nearby meson-meson thresholds, considering the residual interaction between mesons,
 in order to properly describe the masses of the recently found $D_{s0}(2590)^+$. The analysis is done in the framework of the constituent quark model (CQM) proposed in  Ref.~\cite{Vijande:2004he,Valcarce:2005em}. This model successfully describes hadron phenomenology and
reactions~\cite{Fernandez:1992xs, Garcilazo:2001md, Vijande:2004at} and has
recently been applied to (non)conventional hadrons containing heavy quarks~\cite{Segovia:2011zza, Ortega:2012cx, Ortega:2016pgg,
Ortega:2020tng,Ortega:2021zgk}.


\section{Theoretical formalism}
\label{sec:Theory}

For the present analysis, we will use the same theoretical formalism as in the study of $J^P=0^+$ and $1^+$ $D_s$ low-lying spectrum of Ref.~\cite{Ortega:2016mms}.

\subsection{Constituent quark model}
\label{subsec:CQM}

Details of the constituent quark model, explicit expressions for all the potentials and the value of the model
parameters can be found in Ref.~\cite{Vijande:2004he}, updated in
Ref.~\cite{Segovia:2008zz}. Thus, let us briefly discuss the most relevant aspects of the model.

Within the CQM, light quarks acquire a dynamical constituent mass due to the dynamical breaking of the chiral
symmetry in QCD. Such spontaneous chiral symmetry breaking, which induces the exchange of Goldstone bosons between light quarks, can be modelled with the effective Lagrangian~\cite{Diakonov:2002fq}
\begin{equation}
{\mathcal L} = \bar{\psi}(i\, {\slash\!\!\! \partial}
-M(q^{2})U^{\gamma_{5}})\,\psi  \,,
\end{equation}
which is invariant under chiral transformations.
In the previous equation, $M(q^2)$ is the dynamical (constituent) quark mass and $U^{\gamma_5} =
e^{i\lambda _{a}\phi ^{a}\gamma _{5}/f_{\pi}}$ is the matrix of Goldstone-boson
fields. Expanding such matrix, we can write
\begin{equation}
U^{\gamma _{5}} = 1 + \frac{i}{f_{\pi}} \gamma^{5} \lambda^{a} \pi^{a} -
\frac{1}{2f_{\pi}^{2}} \pi^{a} \pi^{a} + \ldots
\end{equation}
where the first term generates the constituent light quark mass, the second
describes a one-boson exchange interaction between quarks and the third term
comes from two-pion exchanges, which has been modelled by means of a scalar-meson exchange potential.
Goldstone-boson exchanges are not allowed for $QQ$ and $Qq$ quark pairs, being $Q=\{c,b\}$ and $q=\{u,d,s\}$, as the chiral symmetry is explicitly broken for those quark pairs.

Besides, the model incorporates a one-gluon exchange potential, generated from the vertex
Lagrangian
\begin{equation}
{\mathcal L}_{qqg} = i\sqrt{4\pi\alpha_{s}} \, \bar{\psi} \gamma_{\mu}
G^{\mu}_{c} \lambda^{c} \psi,
\label{Lqqg}
\end{equation}
where $\lambda^{c}$ are the $SU(3)$ colour matrices, $G^{\mu}_{c}$ is the
gluon field and $\alpha_{s}$ is the strong coupling constant. The scale
dependence of $\alpha_{s}$ can be found in {\it e.g.}
Ref.~\cite{Vijande:2004he}, it allows a consistent description of light, strange
and heavy mesons.
Let us remark here that the pseudo-scalar ($\pi$) and scalar ($\sigma$) mesons can be safely exchanged together with the gluon without any double counting~\cite{Yazaki:1989rh}.

The model is completed with a confining potential, which models the linearly rising attractive potential
proportional to the distance between infinitely heavy quarks that emerges from multi-gluon exchanges.
In order to consider the influence of sea quarks, the confining interaction is screened at larger distances,
to mimic the breaking of the quark-antiquark binding string~\cite{Bali:2005fu},
\begin{equation}
V_{\rm CON}(\vec{r}\,)=\left[-a_{c}(1-e^{-\mu_{c}r})+\Delta \right]
(\vec{\lambda}_{q}^{c}\cdot\vec{\lambda}_{\bar{q}}^{c}) \,,
\label{eq:conf}
\end{equation}
where $a_{c}$ and $\mu_{c}$ are model parameters. This potential
shows a linear behaviour at short distances, with an effective confinement strength,
$\sigma=-a_{c}\,\mu_{c}\,(\vec{\lambda}^{c}_{i}\cdot \vec{\lambda}^{c}_{j})$,
while it becomes constant at large distances.

The meson energies and wave functions are obtained
by solving the Schr\"odinger equation using the Gaussian Expansion
Method~\cite{Hiyama:2003cu} which provides enough accuracy and it simplifies
the subsequent evaluation of the needed matrix elements.


\subsection{Coupled-channels calculation}
\label{subsec:coupledchannel} 

The hadronic wave function of the present coupled-channels calculation will be described as the
combination of $c\bar s$ states and meson-meson channels,
\begin{equation}
| \Psi \rangle = \sum_\alpha c_\alpha | \psi_\alpha \rangle
+ \sum_\beta \chi_\beta(P) |\phi_A \phi_B \beta \rangle,
\label{ec:funonda}
\end{equation}
where
$|\psi_\alpha\rangle$ are $c\bar{s}$ eigenstates of the two-body
Hamiltonian, $\phi_{M}$ are wave functions associated with the $A$ and $B$ 
mesons, $|\phi_A \phi_B \beta \rangle$ is the two meson state with $\beta$ 
quantum numbers coupled to total $J^{PC}$ quantum numbers and $\chi_\beta(P)$ 
is the relative wave function between the two mesons in the molecule.

Two possible sources of interaction emerge from this hadron state. On the one hand,
the two and four quark sectors can be coupled via the creation of a light-quark pair.
On the other hand, the meson-meson channels interact among them through
a residual interaction derived from the
same intrinsic $qq$ interactions described by the CQM. To derive the latter
meson-meson interaction we use the Resonating
Group Method (RGM)~\cite{Wheeler:1937zza,Tang:1978zz} (see, e.g, Ref.~\cite{Ortega:2016mms} for details).

The RGM allows us to describe the strong interaction at the meson level from the constituent $qq$ interactions. The mesons are taken as $q\bar q$ clusters, and the
residual cluster-cluster interaction emerges from the microscopic interaction among constituent quarks.

The wave function of a system composed of two mesons with distinguishable quarks, $A$ and $B$, is given by
\begin{equation}
\langle \vec{p}_{A} \vec{p}_{B} \vec{P} \vec{P}_{\rm c.m.} | \phi_A \phi_B \beta
\rangle = \phi_{A}(\vec{p}_{A}) \phi_{B}(\vec{p}_{B})
\chi_{\beta}(\vec{P}) \,,
\label{eq:wf}
\end{equation}
where, \emph{e.g.}, $\phi_{A}(\vec{p}_{A})$ is the wave function of the meson $A$ with $\vec{p}_{A}$ the relative momentum between its quark and antiquark. The relative motion of the two mesons is taken into account by the wave function $\chi_\beta(\vec{P})$.

Within RGM, a general process $AB\to A'B'$ can be described by means of exchange or direct kernels, whether there are quark exchanges between clusters or not. The direct potential, ${}^{\rm RGM}V_{D}^{\beta\beta '}(\vec{P}',\vec{P}_{i})$, can be written as
\begin{align}
&
{}^{\rm RGM}V_{D}^{\beta\beta '}(\vec{P}',\vec{P}_{i}) = \sum_{i\in A, j\in B} \int d\vec{p}_{A'} d\vec{p}_{B'} d\vec{p}_{A} d\vec{p}_{B} \times \nonumber \\
&
\times \phi_{A}^{\ast}(\vec{p}_{A'}) \phi_{B}^{\ast}(\vec{p}_{B'})
V_{ij}^{\beta\beta '}(\vec{P}',\vec{P}_{i}) \phi_{A'}(\vec{p}_{A}) \phi_{B'}(\vec{p}_{B})  \,.
\label{eq:RGMdir}
\end{align}
where $\beta$ labels the set of quantum numbers needed to uniquely define a certain meson-meson partial wave,
where $V_{ij}^{\beta\beta '}(\vec{P}',\vec{P}_{i})$ are the $qq$ potentials from the CQM and the sum runs over the constituent particles inside each cluster.

The exchange potentials ${}^{\rm RGM}V_{E}^{\beta\beta'}(\vec{P}',\vec{P}_{i})$ represent a natural way to connect meson-meson channels with different quark content such as $D_s\omega$ and $DK^\ast$. It is given by
\begin{align}
{}^{\rm RGM}V_{E}^{\beta\beta'}(\vec{P}',&\vec{P}_{i}) = \int d\vec{p}_{A'}\,
d\vec{p}_{B'}\, d\vec{p}_{A}\, d\vec{p}_{B}\, d\vec{P}\, \phi_{A}^{\ast}(\vec{p}_{A'}) \times \nonumber \\
&
\times  \phi_{B}^{\ast}(\vec{p}_{B'})
V_{ij}^{\beta\beta '}(\vec{P}',\vec{P}) P_{mn} \times \nonumber \\
&
\times \left[\phi_{A'}(\vec{p}_{A}) \phi_{B'}(\vec{p}_{B}) \delta^{(3)}(\vec{P}-\vec{P}_{i}) \right] \,,
\label{eq:RGMexc}
\end{align}
where $P_{mn}$ is an operator that exchanges quarks between clusters.

The coupling between the $c\bar s$ and $D^{(*)}K^{(*)}$
sectors requires the creation of a light quark-antiquark pair. For that purpose,
we will use the $^{3}P_{0}$ transition
operator~\cite{LeYaouanc:1972ae}, the same mechanism employed to describe the
open-flavour meson strong decays. The non-relativistic operator that describes this process is
\begin{equation}
\begin{split}
T =& -\sqrt{3} \, \sum_{\mu,\nu}\int d^{3}\!p_{\mu}d^{3}\!p_{\nu}
\delta^{(3)}(\vec{p}_{\mu}+\vec{p}_{\nu})\frac{g_{s}}{2m_{\mu}}\sqrt{2^{5}\pi}
\,\times \\
&
\times \left[\mathcal{Y}_{1}\left(\frac{\vec{p}_{\mu}-\vec{p}_{\nu}}{2}
\right)\otimes\left(\frac{1}{2}\frac{1}{2}\right)1\right]_{0}a^{\dagger}_{\mu}
(\vec{p}_{\mu})b^{\dagger}_{\nu}(\vec{p}_{\nu}) \,.
\label{eq:Otransition2}
\end{split}
\end{equation}
where $\mu$ $(\nu)$ are the spin, flavour and colour quantum numbers of the
created quark (antiquark). The spin of the $q$ and $\bar q$ is coupled to
one. The ${\cal Y}_{lm}(\vec{p}\,)=p^{l}Y_{lm}(\hat{p})$ is the solid harmonic
defined in function of the spherical harmonic. The strength of the quark-antiquark pair
creation from the vacuum is $\gamma=g_{s}/2m$, $m$ being  the mass of the
created quark (antiquark).

The $^3P_0$ model only depends on one parameter, the dimensionless strength
parameter $\gamma$. The values of such parameter can be constrained through
fits to charmed, charmed-strange, hidden-charm and hidden-bottom sectors.
A global fit to those sectors was performed in Ref.~\cite{Segovia:2012cd},
finding a running of the strength parameter given by
\begin{equation}
\gamma(\mu) = \frac{\gamma_{0}}{\log\left(\frac{\mu}{\mu_{0}}\right)},
\label{eq:fitgamma}
\end{equation}
$\gamma_{0}$ and $\mu_{0}$ being  free parameters, whereas $\mu$ is the reduced
mass of the quark-antiquark in the decaying meson. In this work,
we use the value of $\gamma$ corresponding to the charmed-strange
sector, i.e., $\gamma=0.379$.

From the operator in Eq.~(\ref{eq:Otransition2}), we define the transition
potential $h_{\beta \alpha}(P)$ within the $^{3}P_{0}$ model
as~
\begin{equation}
\langle \phi_{M_1} \phi_{M_2} \beta | T | \psi_\alpha \rangle =
\delta^{(3)}(\vec P_{\rm cm})\,P \, h_{\beta \alpha}(P) \,
e^{-\frac{P^2}{2\Lambda^2}}\,,
\label{Vab}
\end{equation}
where $P$ is the relative momentum of the two-meson state.
In order to soften the $^3P_0$ production vertex at high momenta,
we follow the suggestion of Ref.~\cite{Morel:2002vk} and used a Gaussian-like
momentum-dependent form factor to truncate the vertex,
where $\Lambda=0.84\,{\rm GeV}$ is the value used in this work. 
This value of the cut-off is taken from past studies of the $J^P=0^+$ and $=1^+$
charmed-strange~\cite{Ortega:2016mms} and charmed-bottom sectors~\cite{Ortega:2016pgg}, so no fine tuning
of parameters is employed in the present work.

Following previous studies on XYZ states~\cite{Ortega:2012rs},
incorporating the coupling with charmed-strange states, the coupled-channels
equations can be written as
\begin{equation}
\begin{split}
&
c_\alpha M_\alpha +  \sum_\beta \int h_{\alpha\beta}(P) \chi_\beta(P)P^2 dP = E
c_\alpha\,, \\
&
\sum_{\beta}\int H_{\beta'\beta}(P',P)\chi_{\beta}(P) P^2 dP + \\
&
\hspace{2.50cm} + \sum_\alpha h_{\beta'\alpha}(P') c_\alpha = E
\chi_{\beta'}(P')\,,
\label{ec:Ec-Res}
\end{split}
\end{equation}
where $M_\alpha$ are the masses of the bare $c\bar{s}$ mesons and 
$H_{\beta'\beta}$ is the RGM Hamiltonian for the two-meson states obtained from
the CQM $qq$ interaction. Solving the coupling with the $c\bar{s}$ states, we
end up with a Schr\"odinger-type equation for the meson-meson wave functions,
\begin{equation}
\begin{split}
\sum_{\beta} \int \big( H_{\beta'\beta}(P',P) + &
V^{\rm eff}_{\beta'\beta}(P',P) \big) \times \\
&
\times \chi_{\beta}(P) {P}^2 dP = E \chi_{\beta'}(P'),
\label{ec:Ec1}
\end{split}
\end{equation}
where we have an effective potential which encodes the coupling with the two-quark sector,
\begin{equation}\label{eq:effectiveV}
V^{\rm eff}_{\beta'\beta}(P',P;E)=\sum_{\alpha}\frac{h_{\beta'\alpha}(P')
h_{\alpha\beta}(P)}{E-M_{\alpha}}.
\end {equation}

In order to explore states above and below meson-meson thresholds, the coupled-channels equations of Eq.~\eqref{ec:Ec-Res} are, then, rewritten as a set of coupled Lippmann-Schwinger equations,
\begin{align}
T_{\beta}^{\beta'}(E;p',p) &= V_{\beta}^{\beta'}(p',p) + \sum_{\beta''} \int
dp''\, p^{\prime\prime2}\, V_{\beta''}^{\beta'}(p',p'') \nonumber \\
&
\times \frac{1}{E-{\cal E}_{\beta''}(p^{''})}\, T_{\beta}^{\beta''}(E;p'',p) \,,
\end{align}
where $V_{\beta}^{\beta'}(p',p)$ is the projected potential that contains the sum of direct, exchange and effective potentials of Eqs.~\eqref{eq:RGMdir},~\eqref{eq:RGMexc} and ~\eqref{eq:effectiveV}, respectively. ${\cal E}_{\beta''}(p'')$ is the energy corresponding to a momentum $p''$, which in the nonrelativistic case is
\begin{equation}
{\cal E}_{\beta}(p) = \frac{p^2}{2\mu_{\beta}} + \Delta M_{\beta} \,.
\end{equation}
Herein, $\mu_{\beta}$ is the reduced mass of the meson-meson system corresponding to the channel $\beta$, and $\Delta M_{\beta}$ is the difference between the meson-meson threshold and the lightest one, taken as reference.

Once the $T$-matrix is calculated, we determine the on-shell part which is directly related to the scattering matrix (in the case of non-relativistic kinematics):
\begin{equation}
S_{\beta}^{\beta'} = 1 - 2\pi i
\sqrt{\mu_{\beta}\mu_{\beta'}k_{\beta}k_{\beta'}} \,
T_{\beta}^{\beta'}(E+i0^{+};k_{\beta'},k_{\beta}) \,,
\end{equation}
with $k_{\beta}$ the on-shell momentum for channel $\beta$.
All the potentials and kernels will be analytically continued for complex momenta; this allows us to find the poles of the $T$-matrix in any possible Riemann sheet.

The bare $c\bar s$ amplitudes $c_\alpha$ from Eq.~\eqref{ec:funonda} can be calculated with the generalized eigenvalue problem:

\begin{equation}
 \left(M_{\alpha'}\delta^{\alpha'\alpha}-\Sigma^{\alpha'\alpha}\right)c_\alpha = Ec_\alpha
\end{equation}
with $\Sigma^{\alpha\alpha'}$ the complete mass-shift of the state,

\begin{equation}
 \Sigma^{\alpha\alpha'}=\sum_\beta \int \frac{\phi(p;E)^{\alpha\beta}h(p)^{\beta\alpha'}}{p^2/2\mu-E}p^2dp.
\end{equation}

The $\phi(p;E)^{\alpha\beta}$ function is the $^3P_0$ vertex dressed by the RGM
meson-meson interaction, defined as~\cite{Ortega:2012rs}

\begin{equation}
 \phi(p;E)^{\alpha\beta}=h(p)^{\alpha\beta}-\sum_{\beta'}\int \frac{T_V(p;E)^{\beta\beta'}h(p)^{\beta'\alpha}}{p^2/2\mu-E} p^2dp,
\end{equation}
 where $T^{\beta'\beta}_V(P',P;E)$ is the $T$ matrix calculated excluding the
effective potential of Eq.~\eqref{eq:effectiveV}.

Then, the meson-meson wave function can be expressed as,

\begin{equation}
 \chi_\beta(p)=-\sum_\alpha\frac{c_\alpha\phi_{\alpha\beta}(p;E)}{p^2/2\mu-E}
\end{equation}

Hence, with the usual normalization for the complete wave function of Eq.~\eqref{ec:funonda},

\begin{equation}
\langle \Psi | \Psi \rangle=\sum_\alpha |c_\alpha|^2+\sum_\beta \langle\chi_\beta|\chi_\beta\rangle=1
\end{equation}
we have that the probability for the bare $c\bar s$ channel is ${\cal P}_\alpha=| c_\alpha|^2$, and for the meson-meson channel $\beta$,

\begin{align}
{\cal P}_\beta&=\langle \chi_\beta|\chi_\beta\rangle = \sum_\alpha\int \left|\frac{c_\alpha\phi^{\alpha\beta}(p;E)}{p^2/2\mu-E}\right|^2p^2dp.
\end{align}

\section{Results}
\label{sec:Results}

A first approach to the $J^P=0^-$ $D_s$ spectrum within the CQM, calculated as a two-body $c\bar s$ state, predicts the masses shown in Table~\ref{tab:spectrum}. As it can be seen, the $1^1S_0$ is properly described, just $15$ MeV above the experimental state, whereas the $2^1S_0$ $c\bar s$ state is nearly $140$ MeV above the experimental $D_{s0}(2590)$ mass, not significantly different from the $80$ MeV predicted by Ref.~\cite{Godfrey:1985xj}.
Similar discrepancies where found in the $0^+$ and $1^+$ sectors, where the $D_{s0}(2317)$ and $D_{s1}(2460)$ states have masses below the predictions of naive quark models. 
The relevance of $S$-wave scattering channels for $D_s$ states was already pointed out in Ref.~\cite{PhysRevLett.91.012003}.
In Ref.~\cite{Ortega:2016mms}, such large mass differences were explained as the effect of the coupling with nearby meson-meson thresholds, which lower the masses of the naive $c\bar s$ states. As the $D_{s0}(2590)^+$ is close to open meson-meson channels, it is reasonable to assume that the bare $2^1S_0$ will strongly couple to them, thus reducing its mass.
We want to stress here that this effect on the mass of the $D_{s0}(2590)^+$ is a consequence of the presence of nearby meson-meson thresholds and, hence, it cannot be explicitly absorbed in a redefinition of the model parameters as explained in Ref.~\cite{PhysRevC.77.055206}.

\begin{table}[!t]
\begin{tabular}{ccc}
\hline\hline
 $n^{2S+1}L_J$  & $m_{\rm CQM}$  [MeV] & $m_{\rm exp}$ [MeV] \\ \hline
  $1^1S_0$  & $1983.896$ & $1968.35\pm0.07$ \\
  $2^1S_0$  & $2728.942$ & $2591\pm6\pm7$ \\
  $3^1S_0$  & $3178.464$ &\\
  $4^1S_0$  & $3487.090$ &\\
  $5^1S_0$  & $3704.448$ &\\
  \hline\hline
\end{tabular}
\caption{\label{tab:spectrum} Masses of the $J^P$ $D_s$ mesons predicted by the CQM used in this work. }
\end{table}

In order to describe this state, we will couple the $1^1S_0$ and $2^1S_0$ $c\bar s$ bare mesons to the $D^{(*)}K^{(*)}$, $D_s^*\eta$ and $D_s^{(*)}\omega$ channels, whose thresholds are detailed in Table~\ref{tab:thres}. All of those channels are in a relative $^3P_0$ partial wave in the $J^P=0^-$ sector.~\footnote{There are other possible meson-meson channels in a relative $S$ wave that can be considered, such as the $D_0^* K$ or the $D_s f_0$ channels. We have analyzed their inclusion, but their impact is small in the final results and can be safely neglected.} Further $c\bar s$ states do not contribute to the mass-shift of the $2^1S_0$, so they can be neglected.

The coupled-channels calculation includes two types of interactions: On the one hand, the residual interaction among the meson-meson channels described by the RGM kernels and, on the other hand, the effective interaction due to the coupling with intermediate $c\bar s$ states, mediated by the $^3P_0$ model. The $D^{(*)}K^{(*)}$ can directly coupled to the $c\bar s$ states through the $^3P_0$ mechanism, whereas the 
$c\bar s\to D_s^*\eta$ and $c\bar s\to D_s^{(*)}\omega$ are OZI-suppressed transitions.

\begin{table}[!t]
\begin{tabular}{cc}
\hline\hline
 Channel  & Mass [MeV]  \\ \hline
 $D^*K$ & 2504.20\,MeV \\
 $D_s^*\eta$ & 2659.96\,MeV\\
 $D_s\omega$ & 2750.93\,MeV\\
 $DK^*$ & 2760.85\,MeV\\
 $D_s^*\omega$ & 2894.75\,MeV \\
 $D^*K^*$ & 2902.16\,MeV \\
  \hline\hline
\end{tabular}
\caption{\label{tab:thres} Threshold masses of the channels considered in this work. }
\end{table}

\begin{figure}[!t]
\centering
\includegraphics[width=0.5\textwidth]{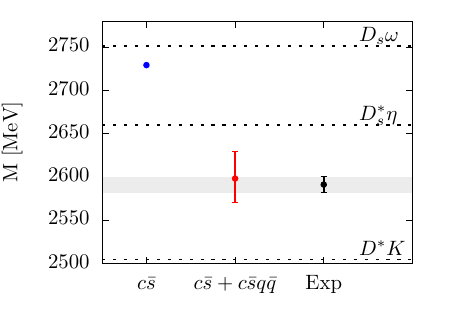}
\caption{\label{fig:masses} Mass of the $D_s(2^1S_0)^+$ state without (blue) and with (red) the effect of coupled channels, compared to the experimental value (black).}
\end{figure}

A first calculation, considering the $D^{(*)}K^{(*)}+D_s^{(*)}\omega+D_s^*\eta$ channels coupled to $1^1S_0$ and $^2S_0$ $c\bar s$, gives us a state with mass $2598$ MeV and a width of $55.5$ MeV for the dressed $2^1S_0$. Here, the parameters of the $^3P_0$ transition are $\gamma=0.38$ and $\Lambda=0.84$ GeV, taken from previous works~\cite{Ortega:2016mms}.  Such mass is just $7$ MeV above the central experimental value of the $D_{s0}(2590)^+$, within the error bar (see Fig.~\ref{fig:masses}), whilst its width is $1.7\sigma$ below.

In order to complete the analysis, we will analyze the stability of the dressed $2^1S_0$ state with the $^3P_0$ parameters $\gamma$ and $\Lambda$. We will allow the $\gamma$ and $\Lambda$ to vary in a 10\% range, that is, $\gamma=0.380\pm0.038$, following Ref.~\cite{Segovia:2012cd}, and $\Lambda=0.840\pm0.084$ GeV. Results are shown in Table~\ref{tab:results}. The sensitivity of the results will be expressed as an error band for the parameters of the state, taken as $\sigma_x=|x_{\rm max}-x_{\rm min}|$.

\begin{table}[h!]
 \begin{center}
 \begin{tabular}{ll}
 \hline\hline
$\Lambda$ [GeV] & $0.840\pm0.084$ \\
$\gamma$ & $0.380\pm0.038$ \\
&\\
$M$ [MeV] & $2598^{+31}_{-28}$ \\
$\Gamma$ [MeV] & $56^{+9}_{-16}$ \\
&\\
${\cal P}_{1S}$ [\%] & $1.3\pm0.4$\\
${\cal P}_{2S}$ [\%] & $46^{+4}_{-5}$\\
${\cal P}_{D^*K}$ [\%] & $44.4^{+5}_{-3}$\\
${\cal P}_{D_s^*\eta}$ [\%] & $0.06\pm1$\\
${\cal P}_{D_s\omega}$ [\%] & $0.011^{+6}_{-5}$\\
${\cal P}_{DK^*}$ [\%] & $4.4^{+2}_{-0.3}$\\
${\cal P}_{D_s^*\omega}$ [\%] & $0.057^{+6}_{-11}$\\
${\cal P}_{D^*K^*}$ [\%] & $3.6\pm0.1$\\
   \hline\hline
 \end{tabular}
 \caption{\label{tab:results} Masses, widths and probabilities of the $2^1S_0$ component for the coupled-channels calculation with different values of $\gamma$ and $\Lambda$, as detailed in the text. The central value of each magnitude is the result with $\gamma=0.38$ and $\Lambda=0.84$ GeV. The error band is calculated as the difference between the maximum and minimum value of each property ($\sigma_x=|x_{\rm max}-x_{\rm min}|$) when the uncertainties of $\gamma$ and $\Lambda$ are considered.}
\end{center}
\end{table}

\begin{figure}[!t]
\centering
\includegraphics[width=0.5\textwidth]{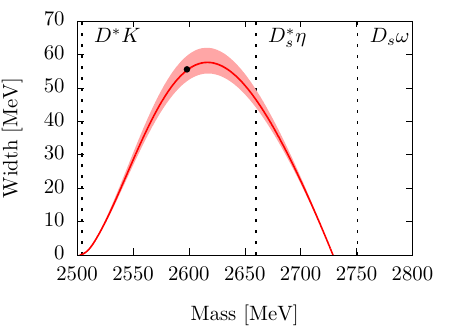}
\caption{\label{fig:traj} Trajectory of the bare $2^1S_0$ $c\bar s$ pole with increasing values of the $^3P_0$ $\gamma$ parameter. The solid red line shows the result for the central cut off value ($\Lambda=0.84$ GeV), whereas the red shadowed area shows the error band from the considered $10\%$ uncertainty for $\Lambda$. The black dot shows the position for $\Lambda=0.84$ GeV and $\gamma=0.38$.}
\end{figure}

The state is an $46\%$ $c\bar s$ state, with a large molecular component, mostly $D^*K$ with a $44.4\%$, which is reasonable as it is the closest channel to the dressed sector. This result is, though, different from the $8\%$ probability of the $D^*K$ component found in Ref.~\cite{Xie:2021dwe}, which could be due to the inclusion of more channels and a more complete description of the interaction between the meson-meson channels done in our work. In Fig.~\ref{fig:traj} we show the evolution of the bare $2^1S_0$ $c\bar s$ pole with increasing coupling to meson-meson channels. We see that the obtained state corresponds to the dressed $c\bar s$ pole and not a new state, even if the probability of meson-meson channels represent about half. We also appreciate that, initially, the state reduces its mass and acquires a large width with increasing coupling strength and, at some point near the central strength value $\gamma=0.38$, the trajectory starts to bend towards the real axis, thus reducing its width, which goes to zero when we approach the $D^*K$ threshold. This pattern is similar to the unitarization effect studied in Ref.~\cite{Hammer:2016prh}.

The effect of the $D_s^{(*)}\omega$ and $D_s^*\eta$ channels is found to be negligible. They are disconnected from the main interaction in this sector, which is the coupling with $c\bar s$ states via the $^3P_0$ mechanism, and their interaction with $D^{(*)}K^{(*)}$ is small as  they are connected through quark rearrangement processes, which are much more suppressed than direct interactions.

The mass of the dressed state is compatible with the $D_{s0}(2590)$ one. The width is still slightly smaller: the upper limit of the theoretical width is $1.2\sigma$ below the experimental one. It is, then, reasonable to wonder if this state is able to describe the experimental line shape of Ref.~\cite{LHCb:2020gnv}, analyzed in the invariant mass spectrum $D^+K^+\pi^-$. Hence, we have calculated the production line shape of the $D^*K+DK^*$ channels, which could contribute to $D^+K^+\pi^-$ through the decay of the $D^*$ and $K^*$, respectively.
 The production amplitude is assumed to be dominated by the $c\bar s$ component of the $D_{s0}(2590)$ wave function,  given by~\cite{Baru:2010ww}

\begin{align}
 \omega_\beta(E) =&  4\pi \mu_\beta k_\beta \left|\frac{\tilde{h}^{\alpha\beta}(k_\beta)}{E-M_\alpha+{\cal G}_{\alpha}(E)}\right|^2 \Theta(E)
\end{align}
with $M_\alpha$ the bare mass of the $2^1S_0$ $c\bar s$ state, $k_\beta$ the onshell momentum of meson-meson channel with $\beta$ quantum numbers, ${\cal G}_{\alpha}(E)$ the exact mass shift of the $c\bar s$ state and $\tilde{h}^{\alpha\beta}(p)$ the $^3P_0$ potential dressed with the RGM interaction,
\begin{equation}\label{ec:tqhmatrix}
 \tilde{h}^{\alpha\beta}(p)=h^{\alpha\beta}(p)-\sum_{\beta'}\int q^2dq \frac{T^{\beta\beta'}(p,q;E)h^{\alpha\beta'}(q)}{q^2/2\mu -E}
\end{equation}

The line shape of the pair $(M_1M_2)_\beta$ can be written as,

\begin{equation}\label{eq:lineshape}
 \frac{d{\rm Br}(M_1M_2)_\beta}{dE} = {\cal A}\,\times \omega_\beta(E)
\end{equation}

We will calculate the sum of $D^*K+DK^*$ line shapes. The only free parameter of the calculation is, then, a global normalization ${\cal A}$, which encodes the $c\bar s$ production vertex amplitude. This normalization can be fitted to the experimental $\{E_i,N_i\pm\sigma_i\}$ points via the $\chi^2$ function:

\begin{equation}
 \chi^2=\sum_{i=1}^N\frac{({\cal A}\, \omega(E_i)-N_i)^2}{\sigma_i^2}
\end{equation}
Actually, the previous $\chi^2$ function can be analytically minimized, giving the normalization:
\begin{equation}
 {\cal A}= \frac{\sum_{i=1}^N \frac{\omega(E_i)N_i}{\sigma_i^2}}{\sum_{j=1}^N \frac{\omega(E_j)^2}{\sigma_j^2}}
\end{equation}

The fit is performed in the range $[2.55,2.90]$ GeV, where the main peak is located. The first peak at $2.53$ GeV is removed, as it corresponds to the $D_{s1}(2536)$ resonance.

\begin{figure}[!t]
\centering
\includegraphics[width=0.5\textwidth]{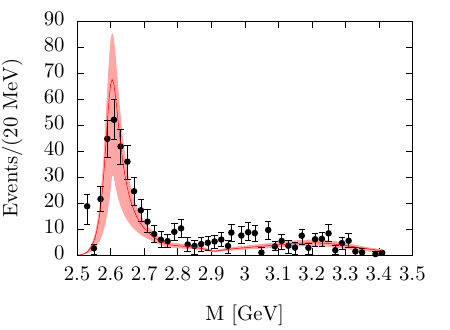}
\caption{\label{fig:lineshape} $D^+K^+\pi^-$ invariant mass spectrum. The colored band shows the $67\%$ C.L.}
\end{figure}

The resulting line shape is shown in Fig.~\ref{fig:lineshape}, with a $\chi^2/{\rm dof}=1.24$. The error band in the figure takes into account the $67\%$ C.L., considering the $\gamma$, $\Lambda$ and ${\cal A}$ uncertainties. As we can see, the theoretical $D_{s0}(2590)^+$ peak agrees with the experimental data.

These results reinforce the assignment of the $D_{s0}(2590)^+$ to the $D_s(2^1S_0)^+$ state, the first radial excitation of the pseudoscalar ground-state $D_s^+$ meson, and puts into manifest the importance of including
the effect of nearby meson-meson thresholds in the description of the masses of mesons.


\begin{acknowledgments}
Work supported by:
EU Horizon2020 research and innovation program, STRONG-2020project, under grant agreement no. 824093;
Ministerio Espa\~nol de Ciencia e Innovaci\'on, grant no. PID2019-107844GB-C22 and PID2019-105439GB-C22/AEI/10.13039/501100011033;
and Junta de Andaluc\'ia, contract nos.\ P18-FRJ-1132 and Operativo FEDER Andaluc\'ia 2014-2020 UHU-1264517.
\end{acknowledgments}



\bibliography{Ds0}

\end{document}